\documentclass[useAMS,usenatbib]{mn2e}
\usepackage{epsfig}

\usepackage{lipsum}
\usepackage{lscape}
\usepackage{longtable,amsmath}

\title[Candidate Massive Clusters]{New Candidate Massive Clusters from 2MASS}

\author[D.\,Froebrich]{Dirk Froebrich$^{1}$\thanks{E-mail: df@star.kent.ac.uk}\\
$^{1}$ Centre for Astrophysics and Planetary Science, University of Kent,
Canterbury, CT2 7NH, United Kingdom }

\begin{document}

\date{Accepted. Received.}

\pagerange{\pageref{firstpage}--\pageref{lastpage}} \pubyear{2012}

\maketitle

\label{firstpage}

\begin{abstract}

Massive stars are important for the evolution of the interstellar medium. The
detailed study of their properties (such as mass loss, rotation, magnetic
fields) is enormously facilitated by samples of these objects in young massive
galactic star clusters. Using 2MASS we have searched for so far unknown
candidates of red supergiant clusters along the Galactic Plane. Utilising deep
high resolution UKIDSS GPS and VISTA VVV data to study colour-magnitude
diagrams, we uncover six new massive cluster candidates in the inner Galaxy. If
spectroscopically confirmed as real clusters, two of them could be part of the
Scutum-Complex. One cluster candidate has a number of potential red supergiant
members comparable to RSGC\,1 and 3. 

Our investigation of UKIDSS data reveals for the first time the main sequence of
the massive cluster RSGC\,2. The stars of the sequence show an increased
projected density at the same position as the known red supergiants in the
cluster and have $E(J-K)=1.6$\,mag. This either indicates an unusual extinction
law along the line of sight or a much lower near infrared extinction to the
cluster than previously estimated in the literature. We suggest that
psf-photometry in UKIDSS images might be able to uncover the main sequence of
other RSGC clusters.

\end{abstract}


\section{Introduction}
\label{intro}

The pivotal role of massive stars in the evolution of galaxies is undisputed.
They emit enormous amounts of ionising radiation, have extreme mass loss
episodes and ultimately undergo supernova or gamma-ray burst explosions that
inject energy, momentum and chemically enriched material into the surroundings.
This feedback shapes the circumjacent interstellar medium and impacts
(positively or negatively) on subsequent star formation. Most massive stars are
formed in a clustered environment and are short lived
\citep{2011A&A...530A.115B}. Thus, the best way to study them is to investigate
young massive clusters. These clusters typically contain larger samples of
massive stars and provide the additional advantage that one can measure their
distances, ages and reddening relatively accurately.

Recent years have seen a flurry of discoveries (or verification) of massive
star clusters in the Galaxy. These include e.g. Westerlund\,1
\citep{2005A&A...434..949C}, three red supergiant clusters (RSGC) in the
Scutum-Complex (RSGC\,1, 2, 3; \citet{2006ApJ...643.1166F};
\citet{2007ApJ...671..781D}; \citet{2009A&A...498..109C}), other objects in
this extended region (e.g. Alicante\,8, \citet{2010A&A...513A..74N}) and even
clusters at the far end of the bar (Mercer\,81, \citet{2012MNRAS.419.1860D}).
These clusters are usually heavily obscured but have the advantage of being
relatively nearby compared to extragalactic systems, hence allowing us to study
individual stars (or systems) in detail. Furthermore, the brightest stars in
these clusters, the red supergiants (RSG), have such huge luminosities ($-11
\le M_K \le -8$, \citet{2006ApJ...643.1166F}) that they are easily visible
accross the entire Galaxy even behind large amounts of extinction.

Here we aim to investigate if there are further unknown clusters of bright RSGs
in the Milky Way that are detectable in the 2\,Micron All Sky Survey (2MASS,
\citet{2006AJ....131.1163S}). In Sect.\,\ref{data} we outline the methods of our
search for cluster candidates. The discovered objects and some of their
properties are discussed in Sect.\,\ref{results}, with particular emphasis on
the main sequence detection in RSGC\,2 in Sect.\,\ref{msrsgc2}. We conclude our
work in Sect.\,\ref{conclusions}.

\section{Data and Analysis Methods}\label{data}

We aim to identify possible clusters and associations of RSG stars in the
Galactic Plane. Their absolute $K$-band magnitudes can range from -8\,mag to
-11\,mag e.g. in RSGC\,1, 2, 3, \citet{2006ApJ...643.1166F},
\citet{2007ApJ...671..781D}, \citet{2009A&A...498..109C}. This means that they
are brighter than $K$\,=\,10\,mag for distances of 15\,kpc and extinction
values of up to $A_K$\,=\,5\,mag.  We hence select all stars from the 2MASS
point source catalogue \citep{2006AJ....131.1163S} which are brighter than
$K$\,=\,10\,mag and that have a photometric quality flag of $Qflg = "AAA"$.
Since all selected stars are bright, the latter criterion is not essential for
our analysis and only some extremely bright objects are removed. This selection
leaves 1.297.713 point sources distributed on the entire sky. 

Most massive star formation in the Milky Way, and hence most massive clusters
with RSGs, are believed to be situated within a few kiloparsec from the centre
of the Galaxy. Hence, we focus our search on the range $240^\circ \le RA \le
285^\circ$ and $-55^\circ \le DEC \le +5^\circ$, which covers everything less
than 35$^\circ$ from the Galactic Centre. This area of the sky includes the
majority of bright $K$-band sources, leaving 992.387 2MASS objects to
investigate.

The known RSGCs mentioned above show a clear overdensity of $K$-band bright
stars of particular colours in an area of several arcminutes in size. We adopt
the near infrared (NIR) colours and magnitudes of the spectroscopically
confirmed RSG stars in RSGC\,3 from \citet{2009A&A...498..109C}. The authors
determine a cluster distance of about 6\,kpc and a NIR extinction of about
$A_K$\,=\,1.5\,mag for the spectroscopically confirmed cluster members. The NIR
colour and magnitude box that encloses most of the RSGs in the cluster has the
borders 2.7\,mag\,$\le (J-K) \le$\,3.8\,mag and 5.0\,mag\,$\le K \le$\,7.0\,mag,
i.e. the box is 2\,mag wide in $K$ and 1.1\,mag wide in $(J-K)$. We can simply
'de-redden' this colour magnitude box to various distances and extinction values
to search for local overdensities (see below). For the purpose of this we adopt
$A_K$\,=\,1.62\,$\times E(J-K)$ following \citet{1990ARA&A..28...37M}, but note
that due to our search strategy the actual choice of extinction law is
completely irrelevant for the cluster candidate detection.

We 'de-redden' the colour magnitude box to distances $d$ between 4\,kpc and
12\,kpc (in steps of 2\,kpc) and extinction values $A_K$ from 1.0\,mag to
2.5\,mag (in steps of 0.25\,mag).  This range of distances and $A_K$ values
should allow us to identify all clusters with RSG stars in the entire inner
Galaxy. Note that even if we identify a cluster candidate for a given $d$ and
$A_K$ combination, it will not necessarily be an object with these parameters
since the apparent magnitudes and colours of the RSG cluster members can vary
quite significantly with spectral type, age and local extinction (see e.g.
Fig.\,19 in \citet{2006ApJ...643.1166F}). For each of the $d$ and $A_K$
combinations we determine a map of local stellar overdensities as described
below. In total there are 35 maps due to the five distances and seven extinction
values choosen.

Of the known clusters with red supergiants, RSGC\,1 seems to be the most
compact with almost all RSGs less than 2\arcmin\ from the cluster centre
\citep{2006ApJ...643.1166F}. In contrast RSGC\,2 has many members that are more
than 4\arcmin\ from the centre \citep{2007ApJ...671..781D}. As a compromise we
search for clusters within a 6\arcmin\,$\times$\,6\arcmin\ box. To determine
the local excess of bright stars in this box, we choose as the control field a
18\arcmin\,$\times$\,18\arcmin\ sized box surrounding the central area. We
apply an oversampling of two, hence the pixelsize is 3\arcmin\ in the final
star density maps.

We estimate the mean star density in the central area for each colour-magnitude
box and subtract the mean star density in the surrounding control field. If
there is an overdensity the pixel value in the final map will be the local
stellar overdensity. Otherwise the pixel value is set to zero to subtract the
large scale stellar density. We then average all the maps for the same distance
and different extinction values, resulting in five maps of local overdensities,
one for each distance. The averaging will increase the signal to noise ratio in
the star density maps, since real clusters will be visible in maps made for
several extinction values. This is caused in part by possible differential
reddening of the cluster stars (up to 1.8\,mag in $A_K$ in e.g. RSGC\,1
\citep{2007ApJ...671..781D}) and the fact that the search boxes overlapp
(0.25\,mag steps in $A_K$ convert to shifts of only 0.15\,mag in $(J-K)$
compared to the 1.1\,mag wide search box).  Finally, a Gaussian smoothing with
9\arcmin\ full width half maximum is applied to the maps prior to the search for
candidate clusters.

\section{Results}\label{results}

\begin{table*}
\centering

\caption{\label{properties} Here we list the names and  positions for our
cluster candidates. The positions coincide with the centre of the cluster in our
star density maps, and might hence deviate from the more accurately determined
centres for the already known clusters. We also list the isochrone
\citep{2001A&A...366..538L} parameters used in the colour magnitude diagrams for
our cluster candidates. Note that the isochrones are not used to determine any
cluster parameters, they are merely for orientation. $^*$For all newly
discovered candidates we assume an age for the isochrone of 16\,Myrs. The column
N$_{RSG}$ lists the sum of all membership probabilities for the RSG candidate
stars in each cluster. For the known clusters we list in brackets the number of
spectroscopically confirmed RSGs. This indicates that the summed up membership
probabilities typically amounts to 60\,\% of the RSG population in the known
clusters.} 

\begin{tabular}{lcccccccc}
Name & $l$ & $b$ & RA (J2000) & DEC (J2000) & Age & d & $A_K$ & N$_{RSG}$ \\ 
& [deg] & [deg] & [deg] & [deg] & [Myr] & [kpc] & [mag] &  \\ \hline
RSGC1 & 25.27 & -0.14 & 279.48 & -6.88  & 12 & 4.0 & 2.1 & 7.8 (14) \\ 
RSGC2 & 26.18 & -0.06 & 279.83 & -6.04 & 12 & 6.0 & 1.0 & 16.1 (26) \\ 
Alicante10 & 28.92 & -0.34 & 281.33 & -3.73  & 20 & 6.0 & 1.4 & 4.1 (8) \\ 
RSGC3  & 29.20 & -0.19 & 281.33 & -3.41 & 16 & 6.0 & 1.2 & 9.6 (16) \\ \hline 
F\,1 & 356.93 & -0.40 & 264.93 & -31.75 & 16$^*$ & 6.0 & 1.4 & 6.0 \\
F\,2 & 11.62 & +0.01 & 272.76 & -18.87  & 16$^*$ & 8.0 & 1.5 & 4.8 \\
F\,3 & 16.58 & +0.36 & 274.91 & -14.34  & 16$^*$ & 9.0 & 1.3 & 12.2 \\
F\,4 & 17.98 & -0.12 & 276.03 & -13.33  & 16$^*$ & 6.0 & 1.6 & 6.6 \\
F\,5 & 21.21 & -0.55 & 277.95 & -10.67 & 16$^*$ & 8.0 & 1.0 & 5.5 \\
F\,6 & 31.00 & +0.03 & 281.95 & -1.70  & 16$^*$ & 6.0 & 1.7 & 4.9 \\
\hline

\end{tabular}
\end{table*}

\subsection{RSGC candidate selection}

In each of the five smoothed local overdensity maps we identified the 10 highest
peaks as potential RSGC candidates. If this includes candidates within 5$^\circ$
from the Galactic Centre (this is especially the case for larger distances, i.e.
fainter stars) we also identified the 10 highest peaks along the rest of the
Galactic Plane outside this 5$^\circ$ radius. Note that close to the
Galactic Centre, holes in GMCs lead to high peaks in the density maps, due to
the overall large star density in this area. These peaks will hence always
appear higher than any real cluster candidate away from the central region.
Hence, many real candidates (even RSGC\,3) would be missed without the above
restriction.

We then merged the individual lists of candidates from the different distance
maps. Objects were considered the same, if they are less than 3\arcmin\ away
from each other in two different maps. Please note that we use the density maps
to determine the cluster coordinates. For consistency, we use these coordinates
even for the known clusters. This explains the slight deviation of the positions
in Table\,\ref{properties} compared to published values for the known clusters.

This resulted in 49 cluster candidates, 15 of which are situated within
5$^\circ$ from the Galactic Centre. All of the candidates are cross-matched
against SIMBAD, the current online version\footnote{\tt
http://www.astro.iag.usp.br/$\sim$wilton/} of the open cluster list based on
\citet{2002A&A...389..871D} and the list of cluster candidates from
\citet{2007MNRAS.374..399F}. We recover 5 known clusters. These are RSGC\,1, 2,
3 (\citet{2006ApJ...643.1166F}, \citet{2007ApJ...671..781D},
\citet{2009A&A...498..109C}) as well as Alicante\,10
(\citet{2012A&A...539A.100G}). The remaining fifth object, identical to
[DB2000]\,54 \citep{2000A&A...359L...9D}, has been identified as spurious by
\citet{2003A&A...408..127D}. We confirm this result and after consideration of
the NIR photometry do not consider this object a cluster (see below). We also
identify some objects close to, but not identical to known (candidate) clusters
(Kharchenko\,3 -- \citet{2002A&A...389..871D}, FSR\,1755 --
\citet{2007MNRAS.374..399F}, Bochum\,14 -- \citet{1975A&AS...20..155M},
Westerlund\,1 -- \citet{1987A&AS...70..311W}).

To verify the nature of the new cluster candidates as potential RSGCs we
compare their NIR colour-colour and colour-magnitude diagrams with the known
RSGCs in our sample. For this purpose we utilise 2MASS data as well as deeper
high resolution photometry from the UKIRT Infrared Deep Sky Survey (UKIDSS -
\citet{2007MNRAS.379.1599L}) Galactic Plane Survey (GPS -
\citet{2008MNRAS.391..136L}) and the VISTA Variables in the Via Lactea survey
(VVV - \citet{2010NewA...15..433M}) data release\,1
\citep{2012A&A...537A.107S}. All 2MASS point sources with a quality flag better
than $Qflg = "CCC"$ are used. From the deeper UKIDSS/VVV data we only utilise
sources fainter than 11\,mag in each of the NIR JHK filters to avoid saturation
effects. Furthermore, only objects with {\tt pstar}\,$\ge 0.999656$, {\tt
merged class}\,=\,(-2 or -1), and {\tt priOrSec}\,=\,0 or {\tt
priOrSec}\,=\,{\tt framesetID} are used. This ensures that only stellar like
objects with good photometry are included in the subsequent analysis (see
\citet{2008MNRAS.391..136L} for details).

We performed  a photometric decontamination (based on
\citet{2007MNRAS.377.1301B} and \citet{2010MNRAS.409.1281F}) to determine
photometric cluster membership probabilities for all stars in each cluster.
This was done for the 2MASS and UKIDSS/VVV data sets seperately. Stars in the
inner 3\arcmin\ around the cluster centre are considered part of the cluster,
while stars between 5\arcmin\ and 9\arcmin\ are taken as control objects. We
applied the method detailed in \citet{2010MNRAS.409.1281F} and used the
15$^{\rm th}$ nearest neighbour in the colour-colour-magnitude space to
establish membership probabilities. We varied this number, as well as the size
of the cluster and control field to verify our results. The exact choice of
these 'free' parameters does, however, not influence our results and
interpretation.

\begin{figure*}
\centering
\includegraphics[width=6cm, angle=-90]{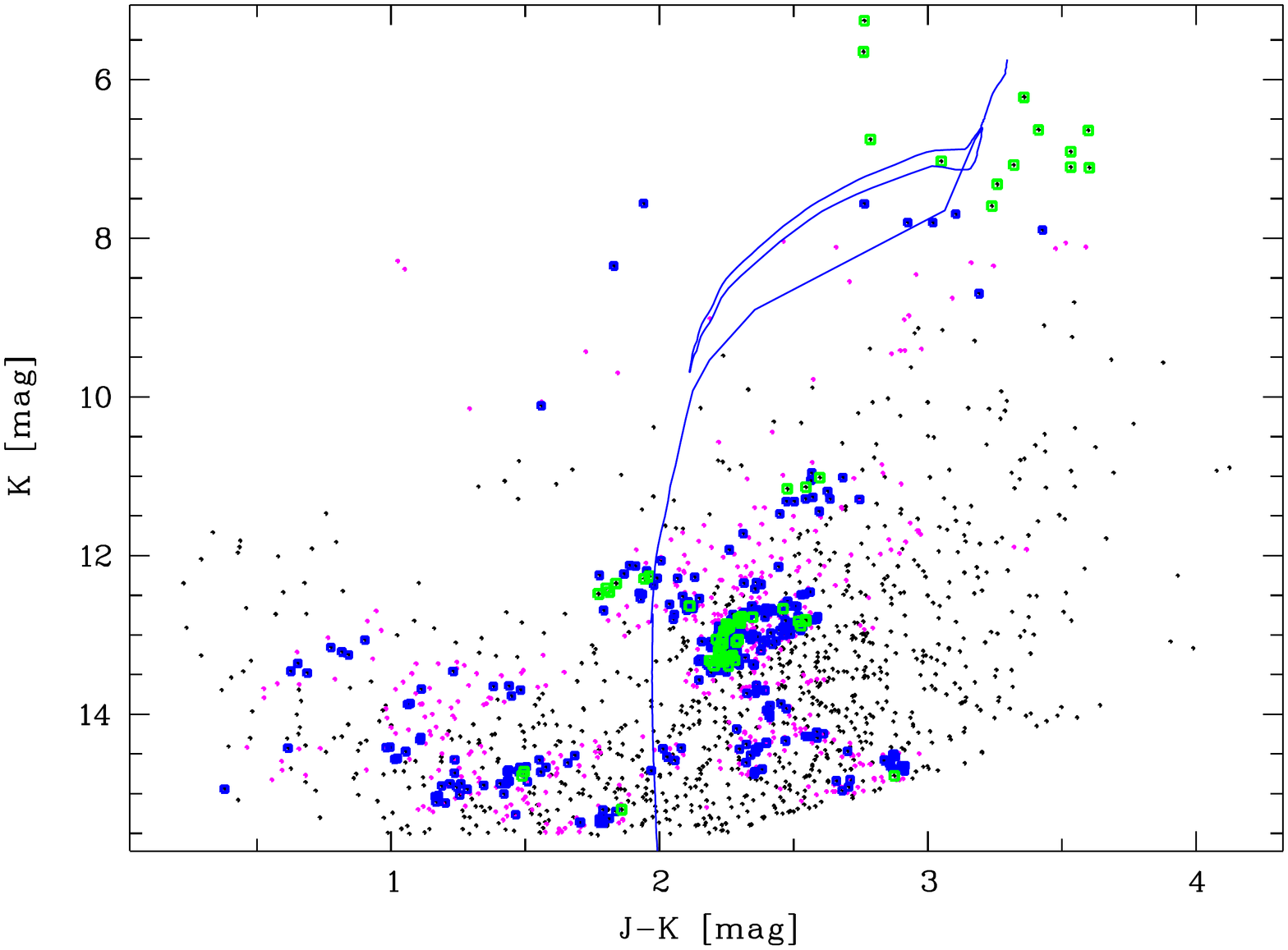} \hfill
\includegraphics[width=6cm, angle=-90]{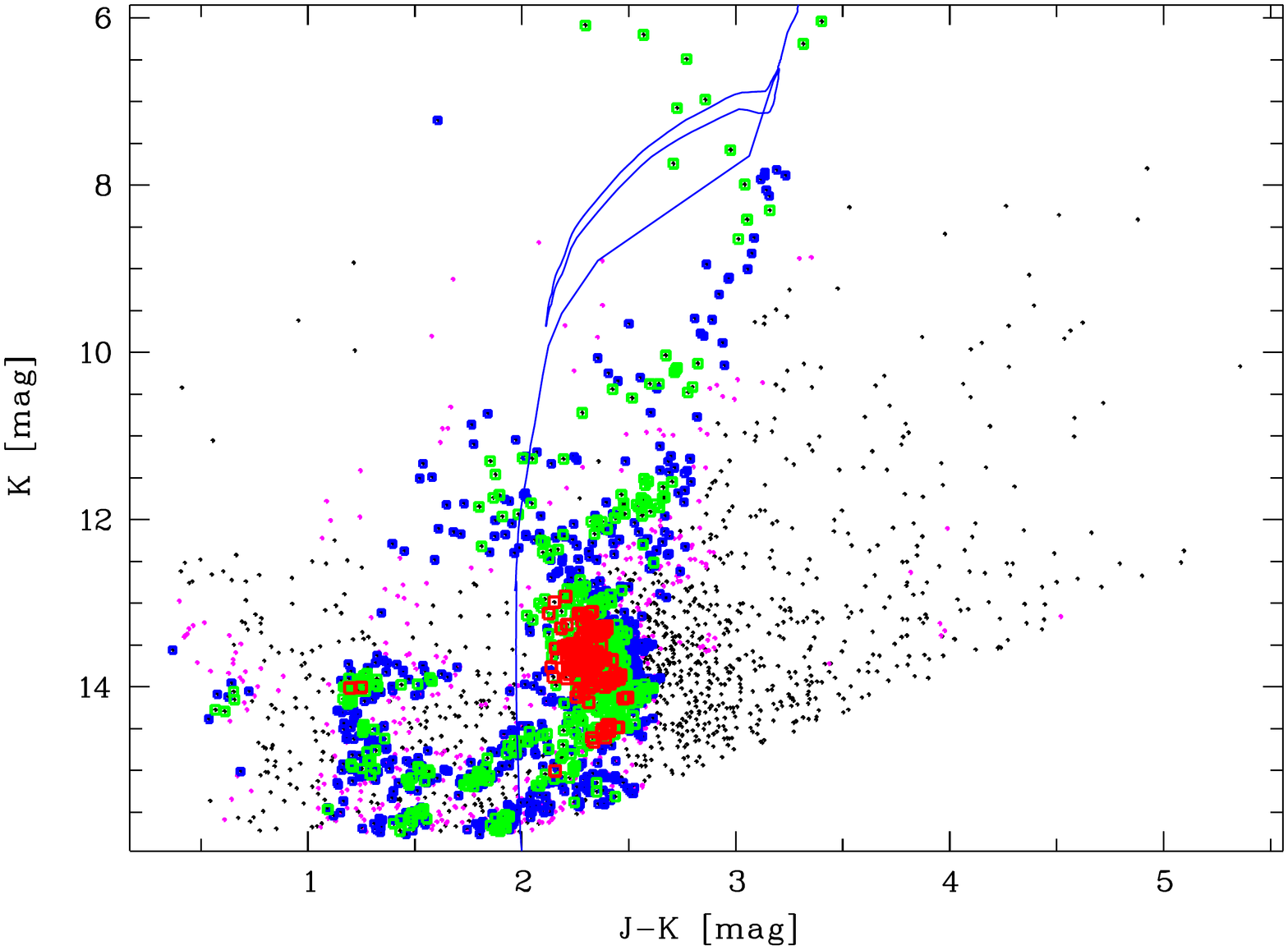}

\caption{\label{f03_cmd}  Combined 2MASS ($K < 12$\,mag) and UKIDSS ($K >
12$\,mag) decontaminated colour-magnitude diagram for cluster candidate F\,3
(left panel) and one of our rejected candidates (right panel). The over plotted
isochrone from \citet{2001A&A...366..538L} has an age of 16\,Myrs, a
distance of 9\,kpc and $A_K$\,=\,1.3\,mag. Larger symbols indicate higher
membership probabilities $P$ with: Red $P>$\,80\,\%, Green
80\,\%\,$>P>$\,60\,\%, Blue 60\,\%\,$>P>$\,40\,\%, Magenta 40\,\%\,$>P>$\,20\,\%
and Black $P<$\,20\,\%.} 

\end{figure*} 

We then inspected all the cluster candidates and selected only objects that
resembled the known RSGCs in our sample. In other words we only selected
candidates that clearly show an overdensity of bright stars (the potential RSGs)
and either no further sign of other potential cluster members (such as for
RSGC\,1) or the indication of a main sequence (such as in RSGC\,2, see below).
Most objects that are removed from our original list are holes in the extinction
in molecular clouds, mimicing cluster candidates. Essentially such holes can
generate sequences of giant stars which resemble the red giant branch of an old
open or globular cluster. As an example we show the combined 2MASS and UKIDSS
colour-magnitude diagram for our cluster candidate F\,3 and one rejected object
in Fig.\,\ref{f03_cmd}. In this plot all stars with $K < 12$\,mag are taken from
2MASS, while all fainter sources are from UKIDSS. Larger symbols indicate higher
membership probabilities and one can clearly see a group of bright red stars
which could be RSGs. In the Table in the Appendix we show the
diagrams for all RSGC candidates and known clusters for comparison. 

After this selection there are 10 RSGC candidates remaining. Four of them are
the known clusters RSGC\,1, 2, 3 and Alicante\,10, while the remaining six are
so far unknown objects. We show photometrically decontaminated diagrams of all
these objects in the Table in the Appendix. For all cluster
candidates we added up the cluster membership probabilities of the RSG
candidate cluster members. Typically these are all stars brighter than $K =
8$\,mag. The sum of the membership probabilities is indicative of the number of
RSG stars (N$_{RSG}$) in the cluster and thus potentially of the cluster mass.
We list these values in Table\,\ref{properties} together with the number of
spectroscopically confirmed RSG stars in each of the known clusters. The number
of confirmed RSG members in the clusters is consistently a factor of 1.6 to 1.8
larger than the value of N$_{RSG}$. 

The colour-magnitude diagrams in the Table in the Appendix
contain an isochrone for each cluster that is based on the Geneva models by
\citet{2001A&A...366..538L}. Table\,\ref{properties} lists the properties
(distance, age, extinction) used for each cluster candidate. Note, that since we
do not know the age of the new cluster candidates, we use 16\,Myrs for all of
them. We do not aim to determine any cluster properties using those isochrones
since it is not possible from photometry alone. The isochrones are only plottet
for orientatio and hence the extinction law used \citep{1990ARA&A..28...37M} to
convert $A_K$ into $E(J-K)$ is not of importance. Furthermore, the applied set
of isochrones from \citet{2001A&A...366..538L} works best with the parameters
listed in Table\,\ref{properties}, which might be slightly different to more
accurate published values for the known clusters in our sample.

\begin{figure*}
\centering
\includegraphics[width=6cm, angle=-90]{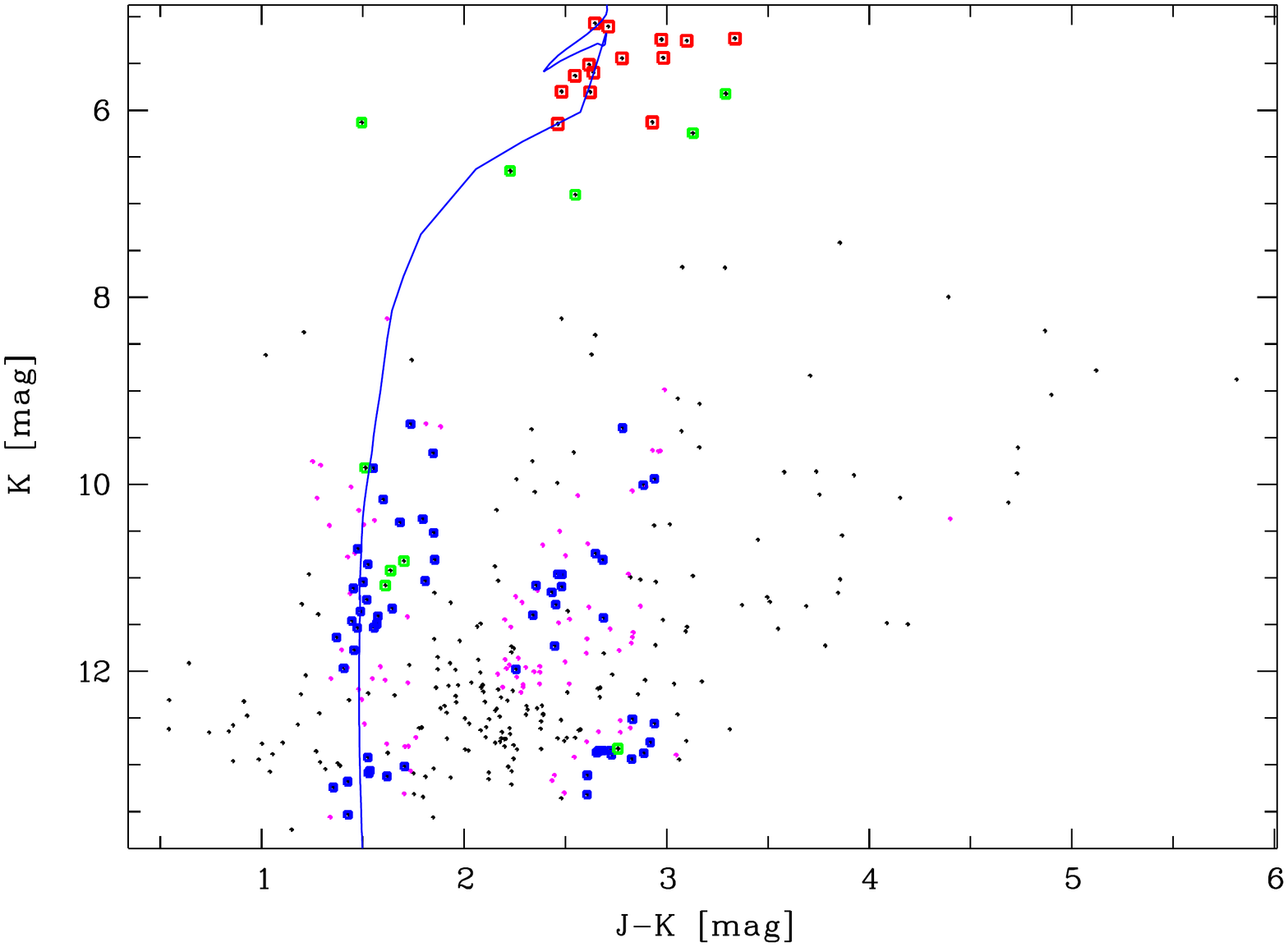} \hfill
\includegraphics[width=6cm, angle=-90]{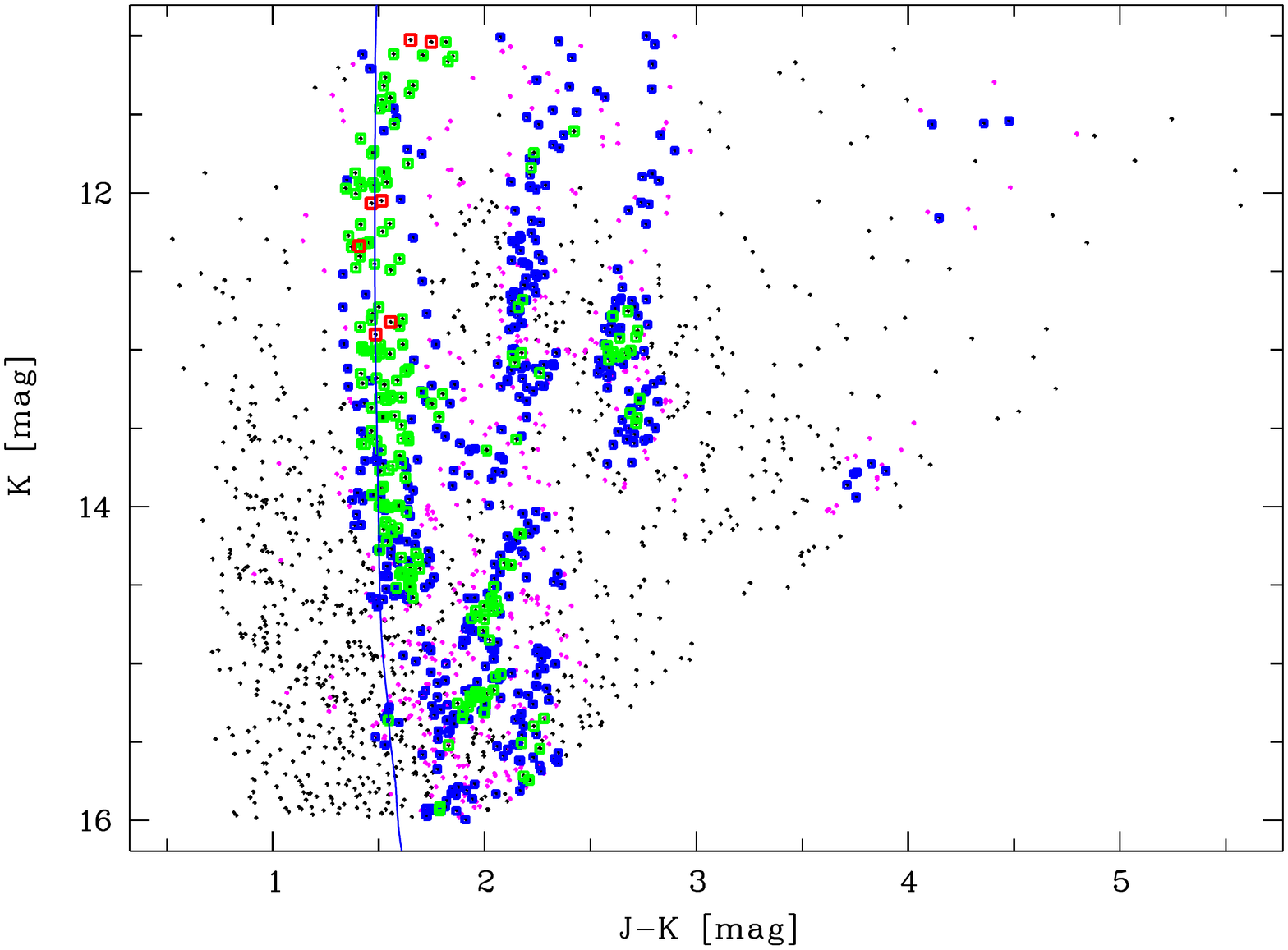}

\caption{\label{f45_cmd} 2MASS (left panel) and UKIDSS (right panel)
decontaminated colour-magnitude diagram for RSGC\,2. The over plotted isochrone
from \citet{2001A&A...366..538L} has an age of 12\,Myrs, a distance of 6\,kpc
and $A_K$\,=\,1.0\,mag. One can clearly identify the main sequence of the
cluster at $(J-K)$\,=\,1.5\,mag. Symbol size and colour coding are the same as
in Fig.\ref{f03_cmd}.} 

\end{figure*} 

\subsection{New Cluster Candidates}

Below we briefly discuss the appearance of the new cluster candidates with
respect to the known objects in our sample.

\begin{itemize}

\item[\bf F\,1:] This is the only candidate that has no deep high resolution
NIR data, as DR\,1 of the VVV contains almost no objects in the cluster area.
However, there are 5\,--\,6 bright stars with extremely similar colours which
could hint that this is indeed a cluster. On the other hand, there is a
sequence of high membership probability objects at $(J-K)$\,=\,3.2\,mag in the
2MASS data, which is indicative of a hole in a cloud. When the next set of VVV
data is released we will have the ability to verify or disregard this object as
a candidate clusters. If object F\,1 is confirmed as real, the sequence of high
probability members at $(J-K)$\,=\,2.1\,mag could be the top of the main
sequence.

\item[\bf F\,2:] This is a weak candidate as it has only a small number of RSG
candidate stars spread over a large range of colours and magnitudes. However,
like RSGC\,1, this object has almost no other high probability cluster members
down to $K = 12$\,mag in the 2MASS data, nor is there any sign of a potential
main sequence visible in the UKIDSS data.

\item[\bf F\,3:] Based on the N$_{RSG}$ value of potential RSGs in the cluster,
this could be the most massive candidate if confirmed. The number of RSG stars
is comparable to RSGC\,2 and 3. The bright cluster members are fainter than for
most of the other candidates, hinting to either a different age or a larger
distance. This is indicated by the distance of 9\,kpc that needs to be used to
fit the isochrone with our assumed age of 16\,Myrs. 

\item[\bf F\,4:] This is a weak candidate as it has a small number of bright
RSG candidate members. However, the sequence of high probability cluster
members at $(J - K)$\,=\,2.5\,mag, visible in 2MASS and UKIDSS data, could be
the upper end of the main sequence. 

\item[\bf F\,5:] Not unlike candidate F\,2, this object has just a handful of
bright high probability members and no other possible members down to the 2MASS
detection limit and there is no clear indication of a main sequence in the
UKIDSS data.

\item[\bf F\,6:] This candidate also just has a small number of potential
members, which are, however, clearly separated (in colour-magnitude space) from
the other stars in the field. 

\end{itemize}

Based on their position, candidates F\,5 and F\,6 could be part of the same
region that is traced by the other massive RSGCs at $25^\circ < l < 30^\circ$.
This Scutum-Complex \citep{2010A&A...513A..74N} is situated at the near end of
the bar which might have triggered an increased star-burst like star formation
activity over the last 20\,Myrs, judging by the inferred ages of the massive
clusters in this region. 

So far only one massive cluster (Mercer\,81, \citet{2012MNRAS.419.1860D},
\citet{2005ApJ...635..560M}) is known to be situated at the far end of the bar
($l \approx 338^\circ$). Our search has not revealed any new candidate clusters
in the area. This could either be caused by a lack of such objects at this
position, or due to observational biases (a combination of larger distance,
increased extinction and crowding). Certainly, searches and analysis of the
deeper, higher resolution VVV images (such as e.g. done by
\citet{2011A&A...532A.131B} and \citet{2012arXiv1206.6104C}), will enable us to
study this region in more detail. We'd like to note that our selection procedure
is optimised to detect cluster candidates similar to RSGC\,1, 2, 3. Other
massive clusters which are younger are not selected. For example Westerlund\,1,
the Arches and Quintuplet clusters are not detected in our maps.

\begin{figure}
\centering
\includegraphics[width=6cm, angle=-90]{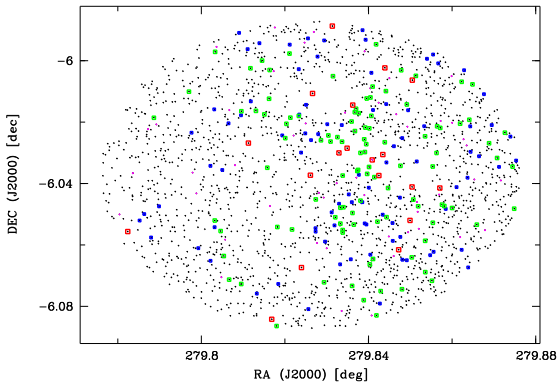}

\caption{\label{special_045} Spatial distribution of high probability cluster
members of RSGC2. Only RSG candidates ($K$\,$<$\,7\,mag) and main sequence
candidates $(J-K)$\,$<$\,1.8\,mag and $K$\,$<$\,14.5\,mag are shown. One can
clearly see that the main sequence objects (mostly green symbols) --  cluster in
the same place as the RSGs (mostly red), slightly off-centre from the adopted
coordinates. Symbol size and colour coding are the same as in
Fig.\ref{f03_cmd}.} 

\end{figure} 

\subsection{The Main Sequence of RSGC\,2}\label{msrsgc2}

With the possible exception of the new objects F\,1 and F\,4, none of the
cluster candidates shows a clear sign of a detection of cluster main sequence
stars. In particular, the known objects RSGC\,1, 3 and Alicante\,10 show
absolutely no indication of potential main sequence clusters members. As can be
seen in Fig.\,\ref{f45_cmd}, our photometric decontamination of the 2MASS and
UKIDSS data reveals a clear sequence of main sequence stars in RSGC\,2. This is
the first detection of the main sequence in any of the known RSG clusters. 

The sequence is less convincing in the 2MASS data, but seems to start at about
$K = 10$\,mag. The small number of high probability member stars fainter than $K
= 11.5$\,mag could be real or caused by increased crowding for these faint
objects. In either case, the sequence is much more obvious in the decontaminated
UKIDSS photometry than in the 2MASS data. We excluded any stars brighter than $K
= 11$\,mag since they may have been saturated. Hence, we cannot determine the
upper end of the main sequence from the UKIDSS data set. As for the 2MASS data,
there is a drop in numbers at the faint end (around $K = 14.6$\,mag) which could
again be attributed to increased crowding for these faint stars. In
Fig.\,\ref{special_045} we show the spacial distribution of the high probability
cluster members, restricted to the RSG stars and the main sequence objects. As
one can clearly see, both groups cluster at the same position, slightly
off-centre in the diagram, a clear indication that the main sequence and the
RSGs are related.

The typical $(J-K)$ colour for the high probability cluster members along the
sequence is $(J-K)$\,=\,1.5\,mag. From the isochrones of
\citet{2001A&A...366..538L} one finds the typical intrinsic colour of massive
main sequence stars is $(J-K)$\,=\,-0.1\,mag. Thus, the cluster main sequence
has $E(J-K)$\,=\,1.6\,mag. \citet{2007ApJ...671..781D} find an average
extinction towards the cluster of $A_K = 1.44 \pm 0.02$\,mag. To bring these
two values into agreement one would need $A_K / E(J-K)$\,=\,0.9. This value is
quite high compared to what is found from other extinction laws e.g. 0.62 from
\citet{1990ARA&A..28...37M}, 0.67 from \citet{2005ApJ...619..931I} or even 0.42
from \citet{2009MNRAS.400..731S}.  If a value of 0.65 is somehow representative
for the extinction law in this region, then $A_K$ for the cluster should be of
the order of $1.05 \pm 0.1$\,mag. This is indeed in agreement with the minimum
(and not the average) extinction values found for the RSG stars in this cluster
by \citet{2007ApJ...671..781D} (see their Table\,2). However, a large fraction
of the RSG stars seem to have a much higher extinction.

Spectroscopic follow up of the potential main sequence cluster members will
allow us to put tighter constraints on the cluster's reddening, age and mass
function as well as the line of sight extinction law. Based on the top of the
main sequence at about $K$\,=\,10\,mag, the age, extinction for the cluster
given in \citet{2007ApJ...671..781D} and the isochrones from
\citet{2001A&A...366..538L}, the most massive stars along the sequence at
$(J-K)$\,=\,1.5\,mag should have masses of about 14\,M$_\odot$ and the sequence
is at least visible down to 5.5\,M$_\odot$ (at $K = 14.5$\,mag).

One question remains: Why can the main sequence only be seen in this clusters
and not, e.g. in the neighbouring clusters RSGC\,1 or 3? In other words: What
sets this object apart from the others? The extinction could be one factor,
since RSGC\,2 has the lowest of all clusters. But since RSGC\,3 has a very
similar $A_K$ value, this can not be the main cause. The abrupt end of the main
sequence about 1.5\,mag above the detection limit (for both data sets, 2MASS and
UKIDSS), hints that increased crowding will effect the photometry sufficiently
to cause difficulties in detecting a main sequence. However, the general star
density in the region of RSGC\,3 is the same as for RSGC\,2 and near RSGC\,1
there are actually only half as many stars per square degree. The most plausible
explanation is that the bright RSG stars in the cluster prevent the detection of
fainter, potential main sequence, stars in their immediate vicinity. This is
evident in the plots of the positions of cluster members in
the Table in the Appendix. As has been mentioned above, RSGC\,2
is the most spatially extended of the known investigated clusters in this paper.
Hence, the bright RSGs do not influence the photometry of such a large fraction
of the cluster area. The compactness of the other clusters might hence prevent a
proper detection of the main sequence stars with aperture photometry. A
re-analysis of the UKIDSS images with psf-photometry should remedy this
situation.

\section{Conclusions}\label{conclusions}

Utilising bright ($K < 10$\,mag) and high signal to noise ($Qflg = AAA$) 2MASS
sources and known colours and magnitudes of RSGs in the massive cluster
RSGC\,3, we have searched for candidate massive clusters along the Galactic
Plane less than $35^\circ$ from the Galactic Centre. Deep NIR photometry from
the UKIDSS and VVV surveys has been used to photometrically decontaminate the
cluster candidate fields to distinguish real cluster candidates from random
associations of bright stars or holes in the extinction of giant molecular
clouds.

As a result of this process we selected 10 good candidate RSG clusters. Four of
these are the known massive clusters RSGC\,1, 2, 3 and Alicante\,10. The
remaining six objects are good candidates for massive clusters in the inner
Galaxy, but need spectroscopic confirmation. Two of the new candidates (F\,5
and F\,6) could be part of the Scutum-Complex, while F\,3 (if confirmed as
real) has a number of RSG candidate stars comparable to RSGC\,2 and 3.

The photometric decontamination of the UKIDSS data of RSGC\,2 reveals for the
first time the main sequence of this cluster which appears to start at about $K
= 10$\,mag. The main sequence stars cluster at the same position as most of the
cluster's RSGs and show $E(J-K)$\,=\,1.6\,mag. This either indicates an unusual
extinction law along this line of sight or a much lower $K$-band extinction
than previously measured for this object (1.05\,mag instead of 1.44\,mag). The
spatial extent of the cluster may have allowed the detection of the main
sequence, in contrast to the other known RSG clusters. Performing
psf-photometry in the UKIDSS images may hence also allow us to detect their
main sequence.

\section*{acknowledgments}

 %
 %

This publication makes use of data products from the Two Micron All Sky Survey,
which is a joint project of the University of Massachusetts and the Infrared
Processing and Analysis Center/California Institute of Technology, funded by the
National Aeronautics and Space Administration and the National Science
Foundation. This research has made use of the SIMBAD database, operated at CDS,
Strasbourg, France. 

\bibliographystyle{mn2e}
\bibliography{references}

\label{lastpage}

\end{document}